\documentclass[acmsmall]{acmart} 

\usepackage{lib/_macros}

\newcounter{highlight}
\newcounter{revised}
\usepackage{makecell} 
\usepackage{multirow} 
\usepackage{csquotes} 
\usepackage[labelformat=simple]{subcaption} 

\usepackage{enumitem}

\AtBeginDocument{%
  \providecommand\BibTeX{{%
    \normalfont B\kern-0.5em{\scshape i\kern-0.25em b}\kern-0.8em\TeX}}}

\setcopyright{none}

\acmConference[CSCW '23]{CSCW '22: The 26th ACM Conference On Computer-Supported Cooperative Work And Social Computing}{October 13-18, 2023}{Minneapolis, MN, USA}
\acmBooktitle{CSCW '23: The 26th ACM Conference On Computer-Supported Cooperative Work And Social Computing, October 13-18, 2023, Minneapolis, MN, USA}

\acmSubmissionID{2791}

\begin{document}

\title{Exploring Immersive Interpersonal Communication via AR}


\author{Kyungjun Lee}
\email{kyungjun@umd.edu}
\orcid{0000-0001-8556-9113}
\affiliation{%
  \institution{University of Maryland}
  \city{College Park}
  \state{MD}
  \country{U.S.A.}
}

\author{Hong Li}
\email{hli8@snapchat.com}
\affiliation{%
  \institution{Snap, Inc.}
  \city{Santa Monica}
  \state{CA}
  \country{U.S.A.}
}

\author{Muhammad Rizky Wellyanto}
\email{wellyan2@illinois.edu}
\affiliation{%
  \institution{University of Illinois at Urbana-Champaign}
  \city{Urbana}
  \state{IL}
  \country{U.S.A.}
}

\author{Yu Jiang Tham}
\email{yujiang@snap.com}
\affiliation{%
  \institution{Snap, Inc.}
  \city{Seattle}
  \state{WA}
  \country{U.S.A.}
}

\author{Andrés Monroy-Hernández}
\email{andresmh@princeton.edu}
\affiliation{%
  \institution{Snap, Inc.}
  \city{Seattle}
  \state{WA}
  \country{U.S.A.}
}
\affiliation{%
  \institution{Princeton University}
  \city{Princeton}
  \state{NJ}
  \country{U.S.A.}
}

\author{Fannie Liu}
\email{fanniefliu@gmail.com}
\affiliation{%
  \institution{Snap, Inc.}
  \city{New York}
  \state{NY}
  \country{U.S.A.}
}

\author{Brian A. Smith}
\authornote{Co-Principal Investigators.}
\email{brian@cs.columbia.edu}
\affiliation{%
  \institution{Snap, Inc.}
  \city{Santa Monica}
  \state{CA}
  \country{U.S.A.}
}
\affiliation{%
  \institution{Columbia University}
  \city{New York}
  \state{NY}
  \country{U.S.A.}
}

\author{Rajan Vaish}
\authornotemark[1]
\email{rvaish@snap.com}
\affiliation{%
  \institution{Snap, Inc.}
  \city{Santa Monica}
  \state{CA}
  \country{U.S.A.}
}

\renewcommand{\shortauthors}{Lee, et al.}


\begin{abstract}
\ifnum\value{highlight}>0
{\color{blue}
\fi
A central challenge of social computing research is to enable people to communicate expressively with each other remotely. Augmented reality has great promise for expressive communication since it enables communication beyond texts and photos and towards immersive experiences rendered in recipients' physical environments. Little research, however, has explored AR's potential for everyday interpersonal communication. In this work, we prototype an AR messaging system, \textit{\name}, to understand people's behaviors and perceptions around communicating with friends via AR messaging. We present our findings under four themes observed from a user study with 24 participants, including the types of immersive messages people choose to send to each other, which factors contribute to a sense of immersiveness, and what concerns arise over this new form of messaging. We discuss important implications of our findings on the design of future immersive communication systems.

\ifnum\value{highlight}>0
}
\fi

\end{abstract}




\begin{teaserfigure}
    \centering
    \includegraphics[width=\textwidth]{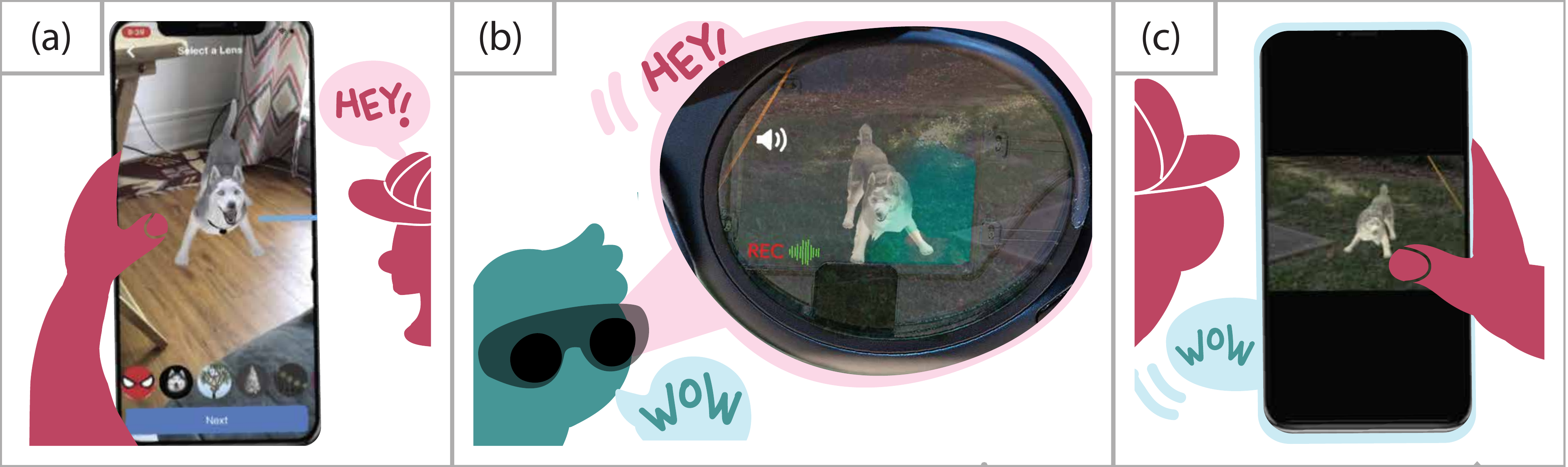}
    \caption{
Overview of \name: 
(a) A smartphone user (sender) sends an AR message (AR content with a voice note) to their friend (recipient) who owns smartglasses. In this case, the AR message is a barking dog accompanied by the sender saying ``Hey!''
(b) \name auto-captures a video of the recipient's reaction (``Wow!'') while they experience the AR message. The video is sent to the sender once the recipient confirms.
(c) The sender views the recipient's reaction video.
    }
    \label{fig:teaser}
    \Description[Three steps, (a), (b), and (c), of using ARwand]{(a) illustrates that the sender adds a virtual dog with a voice note, ``Hey'', to an AR message on a smartphone; (b) shows that the recipient receives the AR message from the sender and experiences the virtual, barking dog in the recipient's environment immersively, while the reaction is being auto-captured; and (c) depicts that the sender watches on the smartphone the reaction video that captures the recipient's reaction to the AR message that shows the virtual dog projected in the recipient's environment.}
\end{teaserfigure}

\maketitle

%

\section{Introduction}




\ifnum\value{highlight}>0
{\color{blue}
\fi

Enabling people to communicate expressively with others remotely is a perennial challenge in social computing research. Technology is integral to facilitating everyday communication, as
people frequently express themselves with texts~\cite{rheingold2007smart, o2014everyday}, photo streams~\cite{gye2007picture, goh2009we, pittman2016social}, and video streams~\cite{ohara2006everyday, bakhshi2016fast}.
However, despite their massive adoption, these existing communication technologies have well-known limitations in conveying important social signals~\cite{mulki2009set}, such as body gestures, touch~\cite{kiesler1984social}, and their inherent physicality~\cite{mennecke2011examination}, which can ultimately affect social presence ~\cite{oh2018systematic} and feelings of connectedness ~\cite{yogarasa2021hivetohive}.

Apostolopoulos and colleagues pitched ``immersive communication'' as a vision for remote communication, involving an ``exchange of natural social signals with remote people... that suspend disbelief in being there'' that could even surpass being face-to-face communication~\cite{apostolopoulos2012road}.
The researchers suggest that advancements in augmented reality (AR) offer a promising direction to bringing this vision closer to reality. They argue that the core research challenge for communication is no longer about how to transmit data quickly enough but rather ``about transduction: how to capture and render information in ways that match the human system, and more fundamentally, what we want our remote experiences to be like.''
Indeed, AR has already shown positive social outcomes that can support this goal, such as social presence, in the areas of learning~\cite{shrestha2022exploring}, collaboration~\cite{osmers2021role, lukosch2015collaboration, guo2019blocks}, and entertainment/play~\cite{marto2022augmented, dagan2022project}.

However, while researchers have demonstrated the value of AR in these specific contexts, less work has explored AR’s potential in informal interpersonal communication. 
Due to the frequency at which billions of people share texts, pictures, and videos to stay in touch, interpersonal communication using the nascent AR technology could lead to an impactful breakthrough.
AR could help people go beyond ``texts'' and ``photos'' to immersive experiences that are more expressive, embedded in their physical context, and feel more like being ``there'' with others, and ultimately more connected with each other.


To explore the potential of AR as an interpersonal communication medium, we built \emph{\name}, named after the ability to magically immerse people in AR with the wave of a ``wand.''
\name enables people to send and react to a new form of message: \emph{AR message}, a piece of AR content with an accompanying voice note rendered in the recipient's physical environment.
For instance, a user can send an AR dog (e.g., a 3D model of a playful dog) into the recipient's view, creating a playful experience by directly augmenting their real-world environment as if the dog were playing near them, as shown in Fig.~\ref{fig:teaser}. They can also schedule the AR dog to appear at a particular time or place---i.e., a trigger condition.
We conducted a user study deploying \name with 24 participants (12 pairs) to understand opportunities and challenges of using AR for immersive casual interpersonal communication.
We present qualitative findings under four emerging themes observed from the study results: (i) What kind of messages do people send with AR messaging, (ii) How do people perceive and use triggers for AR messages, (iii) What makes AR messages immersive, and what concerns arise, and (iv) How and why do people react to AR messages.

We found that senders felt that immersive AR messages could allow them to describe and share objects that are otherwise hard to describe or share, and that they evoke greater emotions in recipients than traditional forms of asynchronous messaging.
Using triggers that enable senders to schedule their messages based on the recipient's context, senders found that their messages were well-consumed by their recipient when the messages appeared in the recipients' environments at the right moment. For successful delivery, they desire to know about the recipients, but we observe that some privacy concerns may arise.
In addition, recipients remarked enjoying AR messages immersively featured in their environments. Still, they mentioned that too much immersion could be troublesome in daily use because it could occlude their view. Nonetheless, recipients desired to share their genuine reactions to the messages with the sender and wanted to have more control over this sharing interaction.

Our paper’s main contributions are: (i) identifying and designing system components that enable immersive interpersonal communication through AR messaging, and (ii) observing user behaviors and preferences to inform the design of future AR systems for interpersonal communication.

\ifnum\value{highlight}>0
}
\fi

\section{Related Work}

\ifnum\value{highlight}>0
{\color{blue}
\fi

\subsection{Toward Immersive Communication}
Apostolopoulos and colleagues proposed that immersiveness is key to revolutionizing remote communication because it can enable more natural experiences and interactions among people communicating and with the environment. By perceiving more natural social signals, such as voice or life-size body movements, people can ``suspend disbelief in \textit{being there}'' with each other and heighten a ``sense of presence'' even when physically apart~\cite{apostolopoulos2012road}.
Indeed, social presence, or the feeling of being there with another person~\cite{oh2018systematic}, is a crucial component of immersive communication, and can improve feelings of satisfaction and connection in communication~\cite{gunawardena1997social, calefato2010communication}.


Researchers have explored the potential to enhance immersiveness and presence by supplementing communication with existing or new social signals.
For example, prior work has explored ``adding back'' missing social cues, such as touch through haptic systems ~\cite{cha2008hugme, steinbach2012haptic}, enhanced gaze \cite{d2021shared, shen2021energy}, and higher quality video to show body gestures \cite{ignatov2021real}. However, these systems tend to require fairly complex setups and customized hardware, which can be difficult to integrate into everyday communication. Others have explored the addition of indirect, contextual signals to communication, such as biosignals~\cite{liu2021significant, hassib2017heartchat, snyder2015moodlight}, activity streams~\cite{griggio2019augmenting, hasan2020coaware}, and location~\cite{dahl2006you} to heighten awareness of what the other person is up to or their environment. Though these signals can support feelings of presence and help people stay in touch, they may not necessarily create a sense of immersion. They only provide indirect cues for people to imagine the other person or their environment, rather than experience cues directly in their  environment.

Researchers have also explored immersive communication through virtual reality (VR), a medium that allows people to join a collaborative virtual environment comparable to physical world objects and events. Users can move around in the virtual space, manipulate virtual objects, and perform other actions (e.g., communicate with other people via text or audio) in a way that creates a feeling of presence in a virtual environment \cite{weiss1998virtual}. But while VR creates immersion through rich virtual environments sharable with remote others~\cite{gunkel2018virtual}, VR environments are purely digital~\cite{smith2018communication}. Thus, they do not leverage physical world everyday communication contexts that can provide significant meaning to communication content~\cite{garten2019measuring}. For instance, one may not communicate the same way when walking through a public park compared to chatting with friends while playing games at home, because their experiences differ in their environments (e.g., being around others in a public space, discussing a game event). Thus, in this paper, we leverage \textit{augmented reality} (AR), where virtual objects and interactions are tied to the physical world, to integrate everyday personal contexts for more immersive communication experiences.

\subsection{AR for Immersive Communication}
AR refers to enhancing or overlaying physical world environments or objects with computer-generated perceptual information such as visual, auditory, and haptic data \cite{furht2011handbook}. AR can create feelings of immersiveness by enabling natural interaction experiences between people when communicating~\cite{apostolopoulos2012road}. Previous research has applied AR to social contexts, including collaboration \cite{osmers2021role, lukosch2015collaboration, guo2019blocks}, games \cite{marto2022augmented, dagan2022project}, museum visiting \cite{jung2016effects}, online learning \cite{shrestha2022exploring},
\ifnum\value{revised}>0
{\color{blue}
\fi
storytelling \cite{bai2015exploring, yilmaz2017using, chen2021scenear},
\ifnum\value{revised}>0
}
\fi
and AR annotation \cite{nassani2015tag}. While these demonstrate the value of AR in areas such as productivity and entertainment, little work has focused on integrating AR in casual communication. Since AR is closely tied to the contexts in which it's experienced, it's unclear how people would use it in their daily social lives.

Some prior research has explored the immersive potential of AR in interpersonal communication. For example, Almeida and colleagues built a system that enabled one user to be present in the other user’s video image. They found their system to be more engaging and foster social presence more, compared to traditional video chats \cite{de2012ar}. Kim and colleagues similarly found that an AR-based communication system, in which remote people could view virtual 3D versions of each other in their physical space, led to greater feelings of being together than a 2D display-based system~\cite{kim2013enhancing}. 
While they examined interpersonal AR communication, their focus primarily was synchronous chatting. However, much of today's communication occurs asynchronously, particularly over mobile messaging~\cite{li2011better, mols2021always}.  
Asynchronous communication has its challenges and paradigms, where message senders may take more time to craft the ideal message, or question the appropriate time or place to deliver them to recipients~\cite{walther2011theories, jones2009context, cao2013connecting}. 
To our knowledge, only a few researchers have built asynchronous AR communication systems, including those that send messages at specified locations~\cite{singh2004augmented}, or enable note posting on top of live video captured at the user's current location~\cite{lin2011nunote}. They primarily focus on system functionality and have not yet explored the social effects and behaviors around asynchronous AR messaging. For example, given virtual content that could be overlaid on the physical world, what kinds of messages would people be interested in creating and sending to others? How does physical context, such as location, affect those messages, including the message content or even delivery to the recipient? How do recipients perceive the messages, and can they enhance a sense of immersion or presence? We explore these questions in the present work by designing \name, an asynchronous AR messaging system, and evaluating its usage through a user study.

\ifnum\value{highlight}>0
}
\fi

\section{ARwand}
\label{sec:arwand}

We implemented \name to explore immersive casual communication.
\name is a system that enables users to asynchronously create and send AR messages to each other, and react to those messages through short videos. 
\name employs both smartglasses (Magic Leap 1, or ML1) and smartphones (iPhone iOS): smartglasses for receiving AR messages and smartphones for authoring them.

We incorporate both devices based on findings from our preliminary prototyping. We found that it is onerous and time-consuming to author AR messages using smartglasses because users need to choose a piece of AR content, choose trigger types, select location triggers using a point and radius on a map, choose a time window for a time trigger, and more.
\ifnum\value{revised}>0
{\color{blue}
\fi
Due to these usability concerns, we limited authoring capabilities to smartphone users. 
While designing for efficient smartglasses interfaces was out of scope for this work, future research should address these limitations to understand users' preferences when they can author messages using AR, irrespective of the device they are on.
\ifnum\value{revised}>0
}
\fi
The smartglasses app is designed to make the process of receiving AR messages seamless and immersive. The smartphone app is designed to make the complex process of authoring AR messages quick and easy.

\name's end-to-end system enables users to do four primary actions: composing, scheduling, playing, and reacting to AR messages.

\begin{figure*}[t]
    \centering
    \includegraphics[width=1\textwidth]{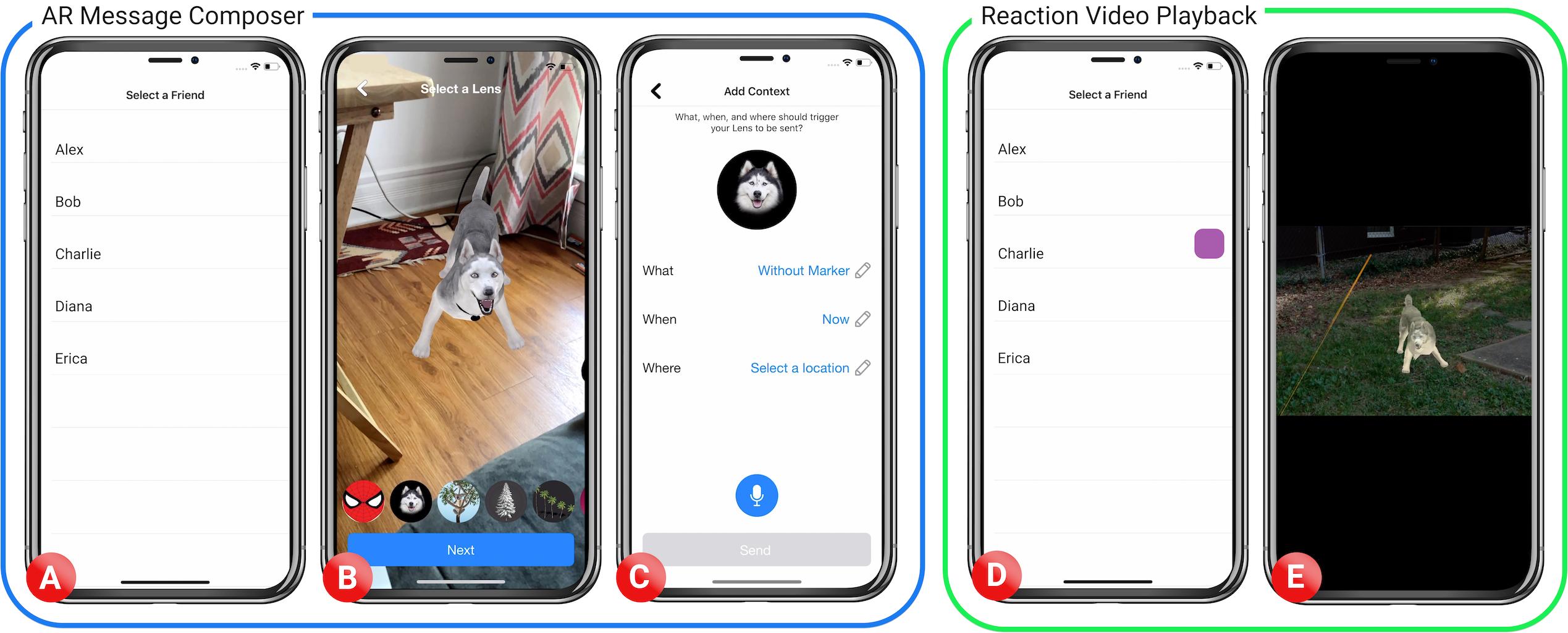}
    \caption{iOS \name app. The user composes AR messages and views the friend's reaction videos on this app. Here we see (A) the ``Select a Friend'' screen, (B) the ``Select AR content'' screen, and (C) the screen for recording a voice note, scheduling the AR message (optional), and sending the AR message. We also see (D) a notification indicating that a new reaction video arrived and (E) the reaction video itself (played by tapping the notification).}
    \label{fig:sender_ios}
    \Description[Three steps, (a), (b), and (c), of composing an AR message on the smartphone, and two steps, (d) and (e), of watching a reaction video on the smartphone]{\name smartphone app: (a) shows a screen where the user can select a friend for AR messaging; (b) shows a screen where the user can choose AR content; and (c) shows a screen where the user can specify message triggers if necessary and record a voice note attached to the message. Reaction Video on the smartphone: (d) shows the screen of showing the friend list, where the user can check if the friend shares a reaction video for the sent AR message; and (e) shows that the reaction video is being played on the smartphone.}
\end{figure*}

\ifnum\value{highlight}>0
{\color{blue}
\fi
\subsection{Composing AR Messages}
\ifnum\value{highlight}>0
}
\fi
The \name iOS app allows users to create and send AR messages.
An AR message is composed of a piece of AR content (such as a virtual object of a dog) and an accompanying voice note (such as ``Hey'').
As shown in Fig.~\ref{fig:sender_ios}, the app consists of three main screens, which we refer to as the following:
(A) ``Select a Friend'' screen, (B) ``Select AR content'' screen, and (C) ``Create voice note and send AR message'' screen.

When the sender opens the app, they are presented with the ``Select a Friend'' screen, which allows them to select a friend (recipient) to send a message to (Fig. \ref{fig:sender_ios}A).
Then, the sender moves onto the ``Select AR content'' screen.
This screen allows the sender to select a piece of AR content to send to their recipient friend (Fig. \ref{fig:sender_ios}B).
After choosing the AR content to send, the sender arrives at ``Create voice note and send AR message'' screen, where they can record a voice note that is up to 10 seconds long by tapping and holding the ``Microphone'' button (Fig. \ref{fig:sender_ios}C).

\begin{figure}[t]
    \centering
    \includegraphics[width=0.5\textwidth]{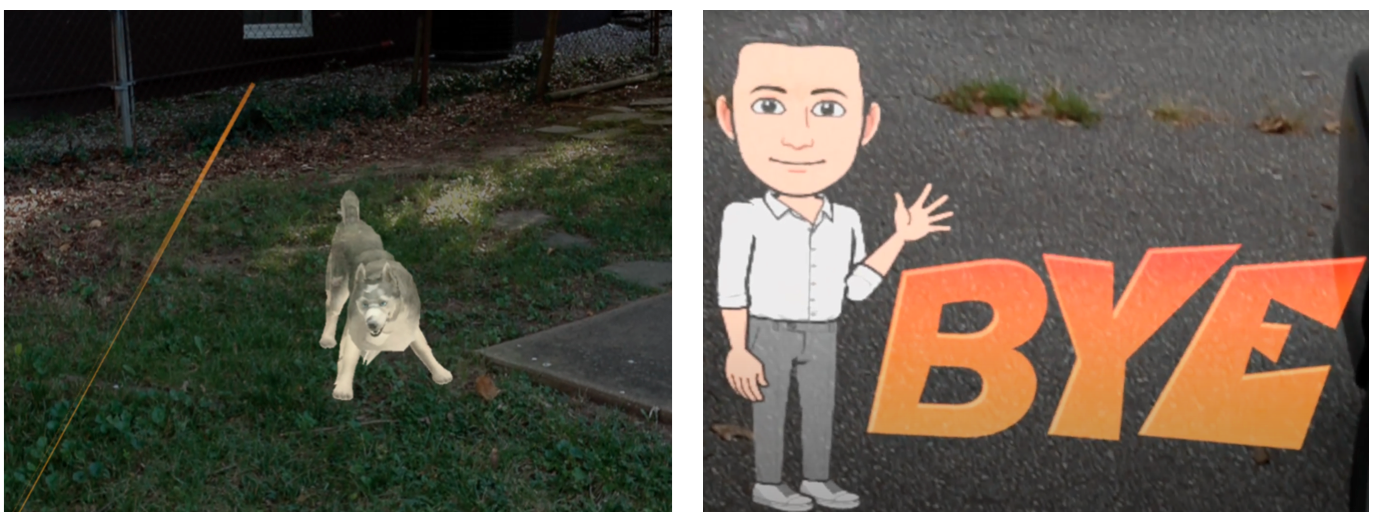}
    \caption{An AR dog and AR avatar as seen through Magic Leap 1.}
    \label{fig:lensAndBitmoji}
    \Description[Two screenshots showing how an virtual dog and an avatar can be viewed, respectively, through the smartglasses (Magic Leap 1)]{The left one shows an virtual dog being placed on the smartglasses user's backyard. The right one shows an avatar with text, ``BYE'', being placed in the recipient's indoor (room) environment.}
\end{figure}

We built \name app to enable senders to add rich, immersive content to their message and explore what kinds of messages they want to express in AR.
We included 21 pieces of AR content that can be rendered in the recipient's environment: 11 virtual objects and 10 avatars (Fig.~\ref{fig:lensAndBitmoji}). 
We selected virtual objects that could represent objects and activities that people might discuss in casual communication, such as food, sports, and nature items (e.g., pizza, basketball, trees and bugs, respectively).
We included avatars as a means for senders to personify themselves in their AR messages. 
Avatars are created via Bitmoji Kit~\cite{bitmoji_kit} as it is a readily available personalized avatar creation tool.
We provide avatars and virtual objects as more immersive, 3D versions of stickers and emojis, which are commonly used in mobile messaging to express one's emotions, fostering feelings of presence, and forming empathy~\cite{wa_stickers_faq,LeeSmiley2016}.
To avoid unrealistic placement of virtual objects, and thus ensure a more immersive and natural experience, they are rendered as either being pinned to the ground or floating in the air. 
For example, a palm tree model is pinned to the ground and a bee model floats in the air as they do in the physical world.
Many of the virtual objects also include associated audio; for example, the bee has its own buzzing sound.
The sender can also adjust the virtual model's rendered scale.
Finally, to add more expressive and personalized meaning to the virtual objects, senders can add voice notes to their AR message.

\begin{figure}
    \centering
    \includegraphics[width=0.95\textwidth]{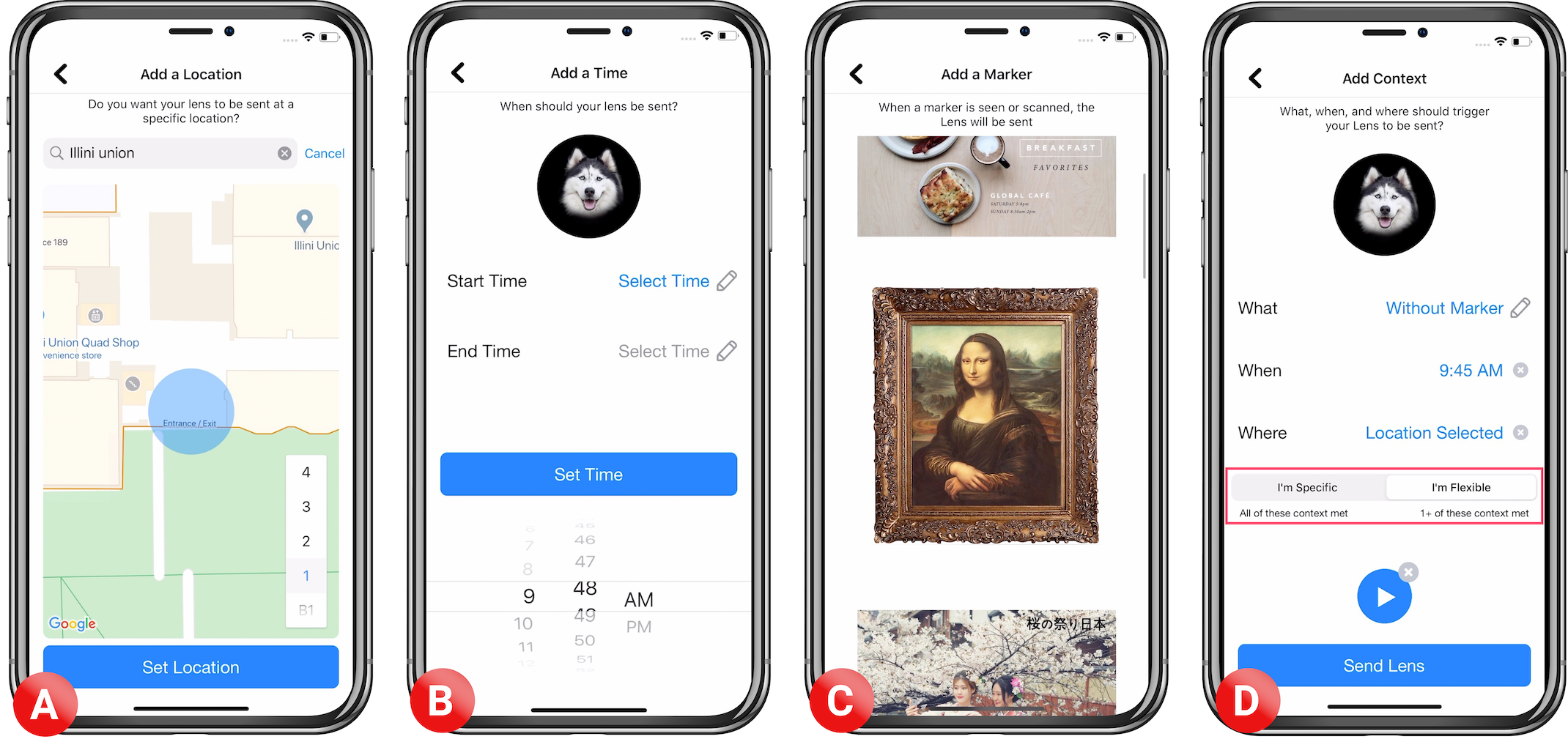}
    \caption{\name scheduler. This is an advanced sender feature that lets them set (A) location, (B) time, and/or (C) visual marker triggers. Senders can also control (D) the triggers' \textit{specificity}. The ``Specific'' option delivers AR messages when \textit{all} triggers are met, while the ``Flexible'' option delivers AR messages when \textit{any} trigger is met.}
    \label{fig:advanced_triggers_screens}
    \Description[Three different types of message trigger, (a) location, (b) time, and (c) visual marker, and the two options of combining multiple triggers (d) on \name scheduler]{(a) shows a map screen where the sender can specify the location to activate an AR message; (b) shows a screen where the sender can specify the time range for message activation (i.e., the start time and the end time of message activation); (c) shows a screen where the sender can specify a visual marker that activates an AR message when the visual marker is within the recipient's view; and (d) shows a screen where the sender can configure the combination of the specified triggers, either ``Specific'' or ``Flexible.''}
\end{figure}

\ifnum\value{highlight}>0
{\color{blue}
\vspace{0.7em}
\fi

\subsection{Scheduling AR Messages}
\ifnum\value{highlight}>0
}
\fi
\name provides senders the option to ``schedule'' their AR message so that the experience is delivered in a specific context based on time, location, or visual markers in the recipient's view.
\ifnum\value{highlight}>0
{\color{blue}
\fi
We implemented this feature due to the importance of real-world context in AR. Prior work that demonstrates the value of using contextual triggers to
relieve cognitive load, improve the sense of presence, and add excitement to messaging ~\cite{jung2005dede, heshmat2020familystories}.
\ifnum\value{highlight}>0
}
\fi
Senders can also choose to send their messages immediately without scheduling them or link them to any context. The system  does not reveal the recipient's location or personal information. Therefore, senders can only schedule AR messages based on their existing knowledge of the recipient's routine and environment. 

\subsubsection{Location trigger}
The location trigger allows senders to trigger AR messages within a specific geographic area (Fig.~\ref{fig:advanced_triggers_screens}A). 
Senders can search for a location by name or zoom to a location on the map, then adjust  
the geofence radius from 7--14 meters (23--46 feet).
\ifnum\value{highlight}>0
{\color{blue}
\fi
We provide location as one of the available contexts because it can help relieve cognitive load for communication that does not require an immediate response. Also, it enables senders to add social meaning to the message~\cite{jung2005dede}.
For example, a sender might want the recipient to see a message about a high school event only at the high school to make the message more relevant and meaningful.
We implemented location selection similar to the mobile-native map app to minimize friction, i.e., a sender can swipe, zoom, and tap to select a location on a map view.
\ifnum\value{highlight}>0
}
\fi


\subsubsection{Time trigger}
The time trigger allows senders to trigger AR messages within a specific time range (Fig.~\ref{fig:advanced_triggers_screens}B). 
The AR message will be delivered successfully if the recipient is wearing their glasses at any point during the specified time range. 
\ifnum\value{highlight}>0
{\color{blue}
\fi
We provide time as one of the available contexts because it can support feelings of presence in asynchronous messaging ~\cite{heshmat2020familystories}.
For example, a sender might want a recipient to view their message during their morning routine, which can help the recipient feel like the sender was communicating with them at that moment, even if they are in different time zones.
We implemented time selection similar to the mobile-native calendar app for ease of use, i.e., a sender can select the hour, minute, and AM/PM options on a time view).
\ifnum\value{highlight}>0
}
\fi


\subsubsection{Visual marker trigger}
The visual marker trigger allows senders to trigger AR messages when the smartglasses detect a specific image (e.g., a poster) in the recipient's view (Fig.~\ref{fig:advanced_triggers_screens}C).
We use the ML1's image tracking algorithm~\cite{ml_image_tracking} to recognize visual markers.
Although we employed this functionality using posters, our implementation is a proxy for a smartglasses system that can recognize physical world objects when smartglasses' onboard computer vision improves.
\ifnum\value{highlight}>0
{\color{blue}
\fi
We implemented this feature because it could enable senders to send a  message when the recipient encounters specific objects.
For example, a sender could compose a ``enjoy your coffee'' message to be delivered only when the recipient's smartglasses detects a coffee shop logo.
\ifnum\value{highlight}>0
}
\fi


\subsubsection{Compound trigger}
\label{sec:compound}
\name also enables the sender to combine multiple triggers.
When a sender selects multiple triggers they see two \textit{specificity option}: ``Specific'' and ``Flexible'' (Fig. \ref{fig:advanced_triggers_screens}D).
If they choose ``Specific,'' the AR message will only be shown if \textit{ALL} the triggers are met (Boolean AND). 
This option lets the sender target a particular context for delivering a message, such as the recipient arriving at a specific park (location) on the weekend (time) and seeing a ``Happy Birthday!'' banner (visual marker) there, but reduces the chance of the AR message being delivered. 
If they choose ``Flexible,''  the AR message will be shown if \textit{ANY} of the triggers are met (Boolean OR).
\ifnum\value{highlight}>0
{\color{blue}
\fi
We implemented this compound trigger to further explore how people balance between specificity and deliverability when scheduling their AR messages.
\ifnum\value{highlight}>0
}
\fi

\ifnum\value{highlight}>0
{\color{blue}
\fi
\subsection{Playing AR Messages}
\ifnum\value{highlight}>0
}
\fi

\name allows recipients to experience AR Messages through a Magic Leap 1 (``ML1'') Unity application.
We chose to implement \name on Magic Leap 1 since it is a commercially available pair of smartglasses that fit \name's design requirements: (i) glasses form factor, (ii) built-in camera, (iii) AR capability, and (iv) wireless Internet connectivity.
When the recipient receives the sender's AR message, a blue flash appears for half a second (as an incoming message notification), followed by the AR message itself (the AR content + the sender's voice note).

\ifnum\value{highlight}>0
{\color{blue}
\vspace{0.7em}
\fi
\subsection{Reacting to AR Messages}
\ifnum\value{highlight}>0
}
\fi
When the recipient receives an AR message, \name captures their reaction by automatically recording their point of view and voice while viewing the experience.
We designed the \emph{reaction video} feature to capture the recipient's instant and authentic reaction to the message, from the moment they start experiencing it. 
Upon the recipient's confirmation, \name will share the reaction video with the sender, allowing the sender to witness the moment themselves and hear what the recipient has to say in response to the experience.
\ifnum\value{highlight}>0
{\color{blue}
\fi
We implemented this reaction recorder to enable the recipient to provide feedback about the AR message, thus completing the communication loop.
Prior work on context-based messaging also highlights the importance of including means for senders to know whether their message was received~\cite{jung2005dede, heshmat2020familystories}.
\ifnum\value{highlight}>0
}
\fi

Reaction videos are composed of (1) the AR content superimposed on the recipient's view, (2) the recipient's voice, and (3) the sender's original voice note, all in the same video recording. 
We designed the reactions in this way to generate a video that can feel like a live recording of their interaction.
For example, if the sender sends a voice message that says, ``look at this puppy,'' along with a virtual dog, and the recipient's voice reaction is, ``wow so cute,'' the final reaction video will appear as if everything happened in the same moment, as all of the 3 components are superimposed.

The recording is 10 seconds long. While recording, a red circle indicator appears on the recipient's view showing a countdown.
Once a reaction video is recorded, the recipient chooses whether to send it to the sender using a ``Yes'' or ``No'' voice command.
If the recipient says ``No,'' the video is discarded.
\name does not save any video or audio data to the device or the cloud.
It is designed to protect the recipient's privacy and give them control over how much information they want to share with the sender. This feature also enabled us to explore when people are willing or not willing to provide instant reactions to AR messages. 
The sender's iOS app notifies them whenever they receive a reaction video.
A purple square also appears next to the recipient's name in the app (Fig. \ref{fig:sender_ios}D).
Tapping the recipient's name plays the reaction video (Fig. \ref{fig:sender_ios}E).


\section{Evaluation}
In this section, we describe the study protocol that we employed in order to study how people use the ability to craft physically situated AR experiences for other people, and how those people would perceive and react to those experiences.

\begin{figure*}
    \centering
    \includegraphics[width=0.7\textwidth]{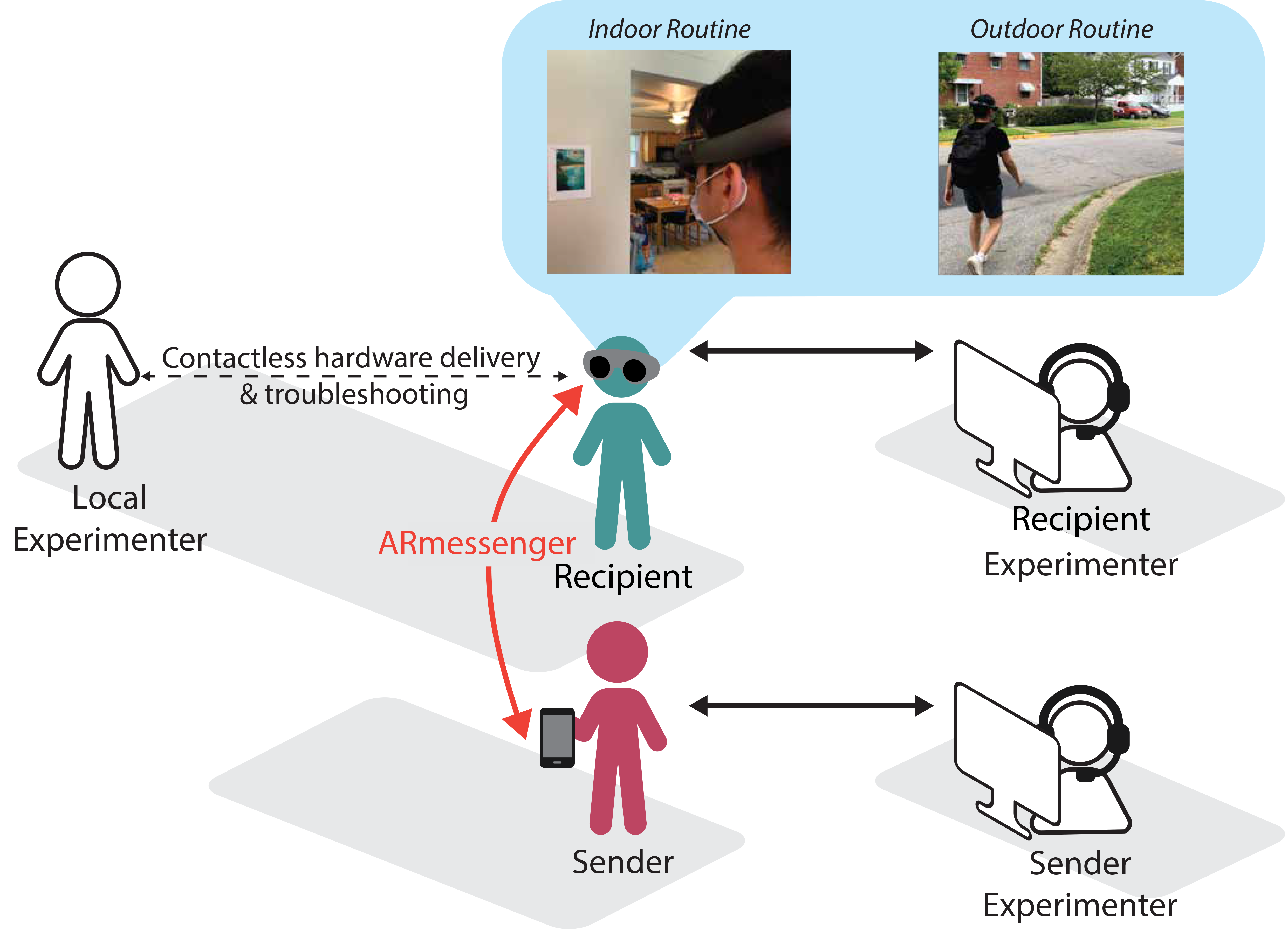}
    \caption{Study setup. Each study involved five people: three experimenters and a pair of participants. The local experimenter and recipient were in the same location, but everyone else was in different locations. The local experimenter made a contactless hardware delivery and provided on-site troubleshooting support, while the recipient experimenter and sender experimenter administered the study. During the study, recipients were asked to perform two routine activities with the Magic Leap 1: spending time at home (Indoor Routine), and walking around their neighborhood (Outdoor Routine).}
    \label{fig:study_diagram}
    \Description[The illustration of our study setup that involves three different experimenters (Local experimenter, recipient experimenter, and sender experimenter) and two participants (recipient and sender)]{Before each study session, the Local experimenter delivers the study equipment to the recipient and stays nearby for troubleshooting support. During the session, the recipient experimenter and the sender experimenter instruct the recipient and the sender, respectively, on a separate video call. The recipient and the sender communicate with each other via ARwand while the sender performs indoor and outdoor routines.}
\end{figure*}

\subsection{COVID-19 Challenges}
The COVID-19 pandemic and \name's reliance on hardware made it impossible for us to employ a traditional in-person study and made the remote study challenging for three key reasons. First, our study involved hardware (Magic Leap 1) that very few people currently own, so we had to make a contactless delivery of the hardware to each participant safely. Second, onboarding and troubleshooting the setup with many moving pieces --- phone, glasses, network, two simultaneous participants --- was non-trivial. However, we communicated frequently with participants alongside having a local experimenter in order to address issues to the best of our ability. Third, most public buildings, college campuses, and stores were closed. Thus, our study simulates a user's daily routine on a walk from their home and neighborhood, described further below.

\subsection{Participants}

We recruited 12 pairs of participants, a total of 24 participants (8 female, 16 male) --- each pair consists of two participants who knew each other. The majority of our participants were undergraduate and graduate students.
Seven were age 18--24, and the rest were age 25--34. Before the study began, participants selected their role as either \textit{sender} (S1--S12), who used the iOS app, or \textit{recipient} (R1--R12), who used the Magic Leap app.

\subsection{Study Setup}
As Fig.~\ref{fig:study_diagram} illustrates, we employed three experimenters for each study session in order to make our study remote and contactless.
A \textit{local experimenter} lived in the participants' city and made a contactless delivery of pre-sanitized study hardware to the participants' front door. The recipient received our ML1 device, but the sender could install our iOS app on their own phone if they wished. The package also included disposable gloves, disinfectant wipes, and hand sanitizer. Two remote experimenters, a \textit{recipient experimenter} and an \textit{sender experimenter}, administered the study via remote video calls with the recipient and sender, respectively. The local experimenter was available to provide on-site troubleshooting during the study when needed.

The study materials also included eight printed visual markers that were placed in and around the recipient's home. \name's scheduling feature included these markers as options for senders to schedule AR messages with.

\subsection{Study Procedure}

\begin{figure}
    \centering
    \includegraphics[width=0.5\textwidth]{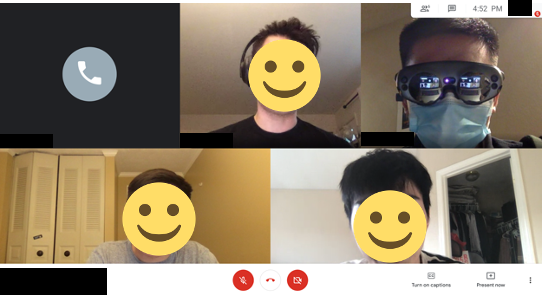}
    \caption{User study in progress. Here, participants share their experiences with \name via a video conference.}
    \label{fig:study_screenshot}
    \Description[The screenshot of a video call]{This screenshot captures a moment where the recipient experimenter and the sender experimenter hear back from the recipient and the sender about their experiences with ARwand at the end of the study session. The Local experimenter also has joined this call in case of troubleshooting.}
\end{figure}

Each study began with the recipient experimenter and sender experimenter joining a video conference with the participant pair.
The participants started by creating two 10-min walking routes in the recipient's neighborhood for the recipient to take. 
Instead of creating a specific scenario for our study, we chose to have participants follow their everyday routine to see how they would create experiences for others during their everyday life.
Next, the sender and recipient tried the end-to-end \name experience out together to get familiar with it. Once they were comfortable, they were split into separate video calls with the sender and recipient experimenter, respectively.

The sender video call consisted of two counterbalanced sessions that took 20--25 minutes each. In one session, the sender sends direct (immediate; non-scheduled) AR messages to the recipient. We asked them to send at least three messages. In the other session, the sender schedules AR messages for the recipient.
We asked them to schedule at least one message per trigger type (location, time, and visual marker).
After each session, participants completed a questionnaire about their experience sending AR messages.

We asked the recipient to wear the ML1 and follow a routine that reflects people's everyday life since the recent outbreak of COVID-19~\cite{whocovid19}, as shown in Fig.~\ref{fig:study_diagram}, where people mostly stay home but also go out for walks in the neighborhood to get some air.
 The recipient was instructed to feel free to \textit{not} share reaction videos with the sender if they did not want to.
At the end of the study, we administered questionnaires and performed semi-structured interviews with both participants to learn about their \name experience (Fig.~\ref{fig:study_screenshot}).
Each study took roughly 2.5 hours in total.

\subsection{Study Analysis}
Our study results include our analyses on participants’ questionnaire responses, qualitative feedback, and usage logs.
The questionnaires included open ended questions and 7-point Likert scales (Strongly disagree--Strongly agree), which we mapped to numerical values (1--7, respectively).
\ifnum\value{revised}>0
{\color{blue}
\fi
We analyzed participants' responses through reflexive thematic analysis~\cite{braun2012thematic}. Three authors generated initial codes using a subset of participants' responses, creating labels based on similarities across the data. The researchers iterated on the codes through several rounds of discussion with two other authors. Then, the initial three researchers applied the finalized codes to the rest of the participants' responses, and used them to generate themes. Finally, all of the authors reviewed and refined the themes together, synthesizing them into major findings on people's perceptions and experiences with \name, which we describe further in the following section.
\ifnum\value{revised}>0
}
\fi


\section{Study Results}

Overall, both senders' and recipients' experiences with \name suggest that AR messages can foster communication and social connection in a uniquely immersive and fun way. 
Our results indicate that senders felt they were indeed augmenting recipients' reality and that recipients felt their reality was being augmented by their friend, therefore creating an immersive experience. 
\ifnum\value{highlight}>0
{\color{blue}
\fi
Participants remarked on this by saying, \textit{``I felt that I was adding some features to my partner's reality which was really cool. [...] The AR messages kind of blend in with the reality''} (S12) and \textit{``I didn’t expect [AR content] to be this clearly pure [...] --- that was really fun.''} (R1).

Participants experienced \name by either sending AR messages and watching reaction videos (for senders) or receiving AR messages and sharing their reaction videos (for recipients).
On average, senders sent 10 AR messages ($Med$=11.5, $s$=3.74) without using any trigger and 9.7 AR messages ($Med$=8.5, $s$=4.21) using triggers. 
Table~\ref{tab:stat_msg_schedule} and Fig.~\ref{fig:msg_delivery} in our appendix present detailed statistics for scheduled messages.

In this section, we provide a more detailed view into participants' usage of \name to communicate based on the emerging themes in our analysis. Put simply, these results reveal the behaviors that emerged from our participants having the ability to send each other AR messages at will. We form our themes into interrogative sentences as follows: (i) What kind of messages do people send with AR messaging? (ii) How do people perceive and use triggers for AR messages? (iii) What makes AR messages and what concerns arise? and (iv) How and why do people react to AR messages?

\subsection{What kind of messages do people send with AR messaging?}

\subsubsection{People create highly expressive messages embedded in the reality}



Participants found that AR messages facilitate uniquely immersive communication experiences.
One participant (S2) remarked, \textit{``We can share objects which are hard to describe in real life so I think it is an augmenting of reality.''}
Senders felt that virtual objects were especially important for heightening immersion compared to existing communication mediums such as text. S2, for example, described the vividness of AR content: \textit{``The messages and objects are hard to be described [with words], but easy to be shown through AR, so it might increase the feeling of reality [over] traditional communicating ways.''} Senders felt that these more immersive messages could evoke greater emotions in recipients than traditional mediums could. S11, for instance, described their decision to position an AR bouncing ball to appear above R11's head, \textit{``so [he] could stop and look up and have the fear that the ball would smash his body[---that] type [of] thing.''}

We found that while the virtual objects made participants' communication experiences more intriguing and expressive, they were not always enough to facilitate communication on their own. Some participants saw the voice note functionality as necessary when crafting AR messages for communication. In general, senders used voice notes to provide meaning to the AR content and to converse through the AR content. R8 acknowledged, \textit{``I think my partner's [voice notes] made the [AR content] I was seeing funny, and not the [AR content] themselves.''}

\subsubsection{People craft messages in a way that is personalized for recipients}

Senders often created messages that reflected their relationship with the recipient, drawing from their shared memories and interests. They used these messages to start new conversations reminiscing on their past experiences together. S7, for example, used the virtual cherry blossom trees to remind R7 of when they \textit{``[went] to [see] cherry blossom[s] in DC.''}
Senders would also tailor their AR message to their partner's interests. For instance, S10 frequently used the virtual basketball object because \textit{``[R10] likes basketball.''} R10 indeed noticed and appreciated the thought behind these personalized messages, stating, \textit{``Any [message] that was tailored to what I like, I like those.''}

In addition, participants sometimes exchanged their inside jokes through AR messages, allowing them to connect in more personal ways. S6 described creating an AR message with the virtual bee, which she scheduled for when R6 went outside for a walk. In particular, S6 remarked, \textit{``I sent [the bee] because we have a backstory[.] When we go for walks, we usually encounter a lot of insects, which I absolutely dislike, and I was like, `This was my payback time!' That was my most favorite one.''}

\subsubsection{People craft stories using the available AR content for inspiration}


We found that senders often considered the AR messages they crafted to be \textit{stories} for recipients to experience rather than simple exchanges of information. In particular, the available set of AR content appeared to serve as inspiration for new story ideas, as opposed to supplementing a story they already had in mind. Senders created their stories by adding voice notes to the AR messages. 
S6, for instance, chose to send a piece of AR representing their avatar holding books, stating, \textit{``There was my avatar which had a pile of books, so I came up with a story that, `Hey, I just graduated with a pile of books!' [...] so it was always me making up stories really.''}

At the same time, some participants expressed that the available content was less conducive to storytelling, particularly those that did not match their physical surroundings. 
S6, for example, did not send any AR messages with the AR dolphin because they found it too incompatible with the environment (since R6 was not near water) and therefore hard to create a story from. They remarked, 
\textit{``I could come up with stories, but like the [AR] dolphin...what could I do? It would never fit into this environment at all.''} 
Thus, we found that, for storytelling, it was important for senders to find AR content that they perceived as compatible with the environment or with their relationship with their friend. 




\subsection{How do people perceive and use triggers for AR messages?}
\label{sec:exp_ar_message_scheduler}



\subsubsection{People use triggers to match AR messages with recipients' physical environments precisely}


We found that senders used \name's trigger feature to ensure that the AR messages they crafted for their partner appeared physically at the intended context---whether that be during a specific activity, at a specific location, or during a specific circumstance. Often, senders found that their ideas for messages or stories to send would be much more effective and well-received if it appeared in the recipients' environments at just the right place or time.
S11, for example, mentioned, \textit{``I could contextualize better and put a tree on the corner wishing him a happy birthday or place a basketball at the end for him to dunk one --- since I know there is a court nearby.''} They knew in advance that R11 (their recipient) would pass by specific locations during the study and crafted AR messages with the location trigger relevant to those locations.

Using the triggers in this way enabled people to craft messages that drew from each other's shared memories and knowledge about each other. These contextual messages thus became more personal and meaningful to participants.
S6 remarked, \textit{``It was a fun experience for me to send the messages [and] bring up stories that the two of us had shared with context rather than sending random messages. You know, things that we have spoken about in the past and things that we know are common.''} Since contextual messages were more targeted, some participants viewed them as special occasion events that recipients would come across, while messages without triggers were more like conversations and agnostic to recipient's environment. S10 said, \textit{``I think the trigger ones...if there was somebody’s birthday or something coming up you could set it for that, so you’re the first one to tell them happy birthday or something. But the ones without the triggers... you’re just kinda sending messages back and forth, versus having to find the message.''}

\subsubsection{Successful triggers require knowledge about recipients}
Participants met varying success when attempting to send messages using triggers, both in terms of crafting and delivering their messages. Although senders sent 9.7 AR messages ($Med$=8.5, $s$=4.21) with triggers on average, recipients received, only 4.9 of the AR messages ($Med$=5.5, $s$=2.54) on average. When sending contextual messages, senders needed to be good at predicting recipients' context. Certain triggers provided specific information that could help them predict context and thus better tailor the message content. 

S12, for instance, could use the recipient's encounters with visual markers as proxies for their activities: \textit{``I think the [visual marker] is the most informative trigger to [the sender] since it kind of narrows down what my partner would do while seeing the [visual marker].''} This participant also noted, \textit{``[The visual marker] trigger in the bathroom means that you’re using the bathroom. Like for sure. And triggers that are in front of the door means that you’re going to get out of the door, like you’re going to get out of the house or something... I like [that the visual marker] really narrows down the partners' location and the action that my partner is doing.''}

While senders could \textit{target} particular activities or particular locations using triggers, they were not always able to successfully predict whether their partners would actually engage in those activities or be at those locations. Their carefully crafted messages would not always be seen, especially when using more specific (e.g., AND mode) triggers. For instance, S10, who scheduled an AR message with the time AND location triggers, mentioned, \textit{``I thought I could anticipate where he'd be on his walk and the time.''} On their own, time triggers offered the most certainty in successful scheduled message delivery since time does not require recipients to do anything to receive the message. We observed, that some senders included a time trigger as an OR condition as a contingency plan for more likely delivery. S6 mentioned combining
\textit{``[multiple triggers with the OR mode] to allow for more flexibility --- for instance, if I missed the location, I was still able to deliver the message based on a time trigger.''}


Since scheduling messages based on triggers required knowing what recipients were doing, some senders were concerned that they might invade recipients' privacy, even though the system itself did not reveal any personal information.
In particular, they mentioned that location-scheduled messages could be intrusive, even when messaging between friends or family members, as they could become aware of the recipient's location upon the delivery of those scheduled messages if the recipient chooses to send a reaction video.
S12 shared, \textit{``I like the trigger system with location, [but what] I didn’t like was [that] a little bit of knowledge of the person’s location is necessary, which I felt is a bit of a privacy violation.''}

\subsection{What makes AR messages immersive, and what concerns arise?}

\subsubsection{Realistic, dynamic content heightens immersion}
Recipients enjoyed the AR messages they received, and found them immersive since they appeared seamlessly and realistically in their physical environments. They felt that realism was achieved through two characteristics. First, they appreciated when AR content
behaved the way they expected in the physical world. For instance, R3 liked the virtual basketball because it bounced like a real basketball, stating, \textit{``I feel there is more interaction when I receive messages that is an animation.''}

Second, recipients enjoyed receiving AR messages that were well-situated in their environment. The messages felt more immersive when the AR content \textit{``appeared in the right place''} (R2), such as an outdoor-related virtual object being \textit{``displayed outdoors''} (R7). Recipients also disliked it when AR content was not placed correctly. R5 shared, \textit{``[an AR message] was initially enjoyable [...] but indoor [virtual] objects were not very well placed.''} At the same time, some realistic AR content could reduce immersion. More specifically, small AR objects (e.g., bees) went unnoticed in situations when a recipient was on the move. On the other hand, when large AR objects (e.g., trees) were rendered too close to recipients' view, it was difficult for recipients to experience them. Echoing this perception, R6 remarked that they were  \textit{``uncomfortable to move my head with the glasses on [to see big virtual objects].''}

\subsubsection{Too much immersion can clash with reality}





Though participants mostly enjoyed the immersiveness of AR messages, they described some situations in which immersion was undesirable. For instance, participants were sometimes shocked when they received messages.
To enable an uninterrupted and seamless experience, ARwand projects the AR message without any notifications or hints. This can add a sense of spontaneity, as if the user comes across the content naturally in their environment; however, recipients felt that the messages could be startling if they did not expect to receive messages.
R6 mentioned, \textit{``There's too much of an element of surprise in receiving the messages [because] I didn't have any control over when I could see the messages.''}
R5 remarked, \textit{``[t]he sudden audio [from AR messages] was something that would startle me.''}

Recipients pointed out that the immersiveness of AR messages could also occlude their view. Blocking the recipient's view can be undesirable, especially when the recipient needs to focus on their visual surroundings. R5 said, \textit{``[one AR message] appeared right at the entry/exit of a parking lot along with sudden music---that could have actually been dangerous.''}
Thus, while received AR messages are rendered immersively in the recipient's environment, they were  too immersive at times.

Finally, while smartglasses enabled greater immersion for AR messages, our findings revealed important social challenges to the form factor. Specifically, participants were sometimes self-conscious when wearing the Magic Leap 1 in public.
As R11 put it, \textit{``It was way too awkward being dressed like Darth Vader [...] with people sitting outside staring at me[.]''}
The social acceptance of wearing AR glasses must be considered as a design factor early on in an AR messaging system or any other system using AR glasses.

\subsection{How and why do people react to AR messages?}

\subsubsection{People react with authentic, live emotional responses}
We observed that recipients reacted to share how they experienced AR messages, ideally wanting to provide similar experiences to the sender. Specifically, they wanted to share \textit{``the context [they] received the message in''} (R5), \textit{``[their] instant reaction to [the message]''} (R6), and \textit{``[their] emotions being aroused''} (R7). Overall, they liked recording their reaction as it allowed them to let their partners (senders) know their genuine responses to AR messages. Participants also wished to reciprocate the messages, so senders could also enjoy the experience. R8 stated, \textit{``I wanted my study partner to be able to experience the AR in a similar and fun fashion[.]''} 

Recipients also appreciated how reaction videos felt like a live recording of their interaction with the sender. R7 mentioned, for example, \textit{``I like how AR [content] and realistic scenes could collapse into one video, which feels immersive.''} R6 wished that the reaction video concept went even further, expressing a desire to \textit{``send back [their] own AR content''} in their reaction videos. Senders also enjoyed viewing recipients' reactions to their AR messages. S11 described, \textit{``It was really funny when he did get [AR messages] that I sent [, and seeing] the reaction... was a really fun experience for me[.] From an entertainment point of view, [I would] use with family and friends[.] If it is something at a reasonable cost, it is something that I would want to have in my life.''}

\subsubsection{People react to converse as if they are physically with each other}



We found that participants viewed the reaction feature as a way to develop conversations with each other. Participants communicated through AR messages and reaction videos similar to how they communicate through traditional messaging systems. R11 remarked on this aspect, \textit{``[As I] received more [AR messages], there was a flow to the communication.''}
Senders also felt that reactions were naturally a source for continuing the conversations. S11 mentioned, \textit{``I liked that [R11] had to say things for me [in a reaction video], [which] choose the next response.''}

Senders further reported that watching reaction videos was enjoyable and made them feel as if they were together with their partner. S4 remarked, \textit{``I [imagined] myself walking and talking with my partner [while crafting my experience].''} The video reactions provided rich feedback to the senders, allowing them to \textit{``see what [the recipient] was seeing and hear [their] response at the same time''} (S4), ``\textit{[look] at the [AR message] I s[ent] and [see] how they perceived it''} (S9), and \textit{``watch the reaction to the [AR content] and also see what she's seeing''} (S12). Thus, viewing how recipients experienced their message from their first-person perspective helped senders feel as if they were actually there with them and sharing the experience together.

While participants conversed in rich ways through \name, they were not always satisfied with the reaction videos. Some participants felt that they were too short, where recipients \textit{``couldn't complete [their] thought more than once [due to the 10-second length limit]''} (S3).
In addition, recipients desired more control over the length of a reaction video and over initiating the recording, mainly due to privacy concerns. R6 remarked, \textit{``[I] couldn't control what was in the background since the response started recording immediately.''} Since the system allowed recipients to manually cancel the transmission of the reaction videos, some recipients chose not to share it in these cases. R1, for instance, recalled not sharing because some private information on their laptop screen was visible in the reaction recording. R7 also mentioned feeling uncomfortable sharing their reaction because \textit{``the[ir] surrounding [...] area is [messy], or less organized.''}

\ifnum\value{highlight}>0
}
\fi

\section{Discussion and Future Directions}
\label{sec:implications}
Our goal with this research was to explore the potential and behavior patterns that emerge from a communication system that enables users to craft immersive experiences for each other, uncovering the design factors and potential pitfalls of such a system in the process. 
Showing our key takeaways in Table~\ref{tab:key_takeaways}, we discuss how our findings can inform the design of future AR messaging systems.

\begin{table}[t]
\ifnum\value{highlight}>0
{\color{blue}
\fi
\renewcommand{\arraystretch}{1.4}
    \centering
    \small
    \caption{
        \ifnum\value{highlight}>0
        {\color{blue}
        \fi
    Key takeaways of design implications derived from our study findings.
        \ifnum\value{highlight}>0
        }
        \fi
    }
    \begin{tabular}{p{0.2\textwidth} p{0.38\textwidth} p{0.38\textwidth}}
        \toprule
        \textbf{Themes} & \textbf{Design Implications} & \textbf{Study Findings} \\
        \midrule
        \multirow{3}{\linewidth}{
        \textit{What kind of messages do people send in AR communication?}} & Support immersive storytelling in the physical world & People crafted experiences that recalled shared memories and created funny moments. \\
            & Explore better content creation tools for everyday, in-the-moment use. & The limited variety of AR content restricted senders' options. \\
            & Embed additional signals for a multi-sensorial communication experience. & People found that AR enabled them to describe objects that are hard to do via other media. \\
        \midrule
        \multirow{3}{\linewidth}{
        \textit{How do people perceive and use triggers for AR messages?}} & Understand how much information recipients are willing to share with others. & People personalized their AR messages using triggers, but doing so had privacy implications. \\
            & Give senders greater control over how to personalize immersive messages for recipients. & People used multiple triggers to increase their scheduled messages' deliverability, but some of them still did not trigger. \\
        \midrule
        \textit{What makes AR messages immersive, and what concerns arise?} &  Make communication immersive without startling or taking over. & People were sometimes startled by the sudden projection of AR experiences \\
        \midrule
        \textit{How and why do people react to AR messages?} & Achieve both authenticity and control. & People continued the story or conversation via a reaction video and wanted more control over it. \\
        \bottomrule
    \end{tabular}
    \label{tab:key_takeaways}
    \ifnum\value{highlight}>0
    }
    \fi
\end{table}

\ifnum\value{highlight}>0
{\color{blue}
\fi

\subsection{Understanding what people want from immersive communication}
Although augmented reality has been shown to benefit productivity~\cite{henderson2007augmented}, education~\cite{radu2019what, villanueva2020meta, hoang2017augmented}, and other applications by virtue of the immersiveness that it provides, our work aims to explore the role that immersive AR can play in the context of informal communication between friends. Specifically, we respond to Apostolopoulos and colleagues' call for communication that can ``capture and render information in ways that match the human system and more fundamentally, what we want our [...] experiences to be like''~\cite{apostolopoulos2012road}. Our studies revealed people's behaviors and preferences around the power to create and experience immersive and contextual experiences \textit{within the physical world} on demand.

Our findings align with what Apostolopoulos and colleagues had hypothesized: highly immersive communication can indeed lead to a better sense of presence and social connectedness. In addition, our results show that, for communication between friends, people use the immersive ability of AR to \emph{craft} and \emph{tell stories} in personal ways for their friends rather than simply to exchange information. We likewise found that recipients highly value when incoming immersive communication is personalized for them both in terms of the content itself and how it is situated and triggered within their environment (i.e., its context). Last, our results reveal many unexpected consequences of immersive communication and subsequent design issues, such as the sudden projection of AR content can startle recipients and block their physical view.
Based on our learning, we share the following design implications.

\subsubsection{Supporting immersive storytelling in the physical world}
People were excited to use immersive AR content creatively to express stories to their friends; thus, future researchers should explore tools that can support the story-crafting and viewing process. For instance, our findings demonstrate the importance of the story's contextual relevance, both in terms of timing and how the message was registered in the physical environment, which heightened recipients' sense of immersion. Future systems can surface locations or art pieces of mutual significance to help create stories associated with that context. 

Additionally, ARwand's AR content asset inventory fell short for some, thus future work should explore better content creation tools for everyday, in-the-moment use to empower people to augment others' environments at will. Future systems can allow people to ``clone'' objects around them to have an unlimited supply of immersive content, create 3D artifacts using paint brush like tools, or connect with online repositories such as SketchFab~\cite{sketchfab}, or integrate with AI text-to-image systems such as DALL-E-2~\cite{ramesh2022hierarchical}. Equipping users with tools to create or access to AR content will allow them to tell any story or express in any way they want. 

Finally, our current approach involved sending a single piece of immersive content at a time and only focussed on visual- and audio-based experiences. For richer and more immersive storytelling, future work can embed additional signals for a multi-sensorial communication experience~\cite{sleepnoremore, vangoghexpo, artnet}. In addition to crafting entire AR scenes with several pieces of content, designers could build on top of existing work on haptic experiences~\cite{kratz2016thermotouch, peiris2017thermovr, liu2021thermocaress} and olfactory research~\cite{amores2017essence, araki2015artificial, bodnar2004aroma}.  Future systems might help a sender transmit various food items on the recipient's dining table, where each AR food item comes with its smell and thermal haptic information, such as an AR pizza which smells like a pizza and feels hot.

\subsubsection{Making communication immersive without startling or taking over}
Since a sudden projection of received immersive content can startle people, future designers and researchers of immersive communication systems should explore how to seamlessly introduce the content to recipients. For example, integrating subtle notifications, such as showing a warning or providing an option to play content at will, could help prepare recipients to view the message. Likewise, since immersive content can blend with users' environment and be perceived as taking over their environment, future work can experiment with translucency features and different sizes/amounts of allowable immersive content. At the same time, researchers should explore \textit{when} and \textit{what kind} of content is appropriate to show users. Our results suggest that immersing people can be undesirable or even dangerous in some situations (e.g., driving). This suggests the need for not only providing means to reject or intelligently ``snooze'' content delivery (e.g., by detecting a moving car or if the recipient is in the middle of a conversation), but also enabling recipients to manually classify the situations in which they do not want to receive an AR message or certain content~\cite{mark2008thecost, srinivas2016designing}. Future work should focus on understanding recipients' preferences, and embedding those preferences in the senders' message crafting process or in the system delivery to avoid startling people or taking over their environment in undesirable ways.

\subsection{Understanding how reactions should work in immersive communication}
The ability to convey reactions and other forms of feedback to messages are central to communication systems~\cite{shannon1948mathematical}. In fact, communication systems often feature a facility for recipients to quickly react to messages---for example, by long-pressing a text message to add a thumbs up or heart reaction to it---as a lightweight way to express how they feel or acknowledge receipt~\cite{hayes2016one}. However, in the realm of immersive communication, these quick forms of reactions may not completely or authentically capture how the recipient experienced or felt when receiving the message. 

Our findings reveal insights about how the concept of message reactions should manifest in immersive communication. In traditional forms of communication, messages are ``what you see is what you get'' (WYSIWIG) because the message will appear to the recipient. Instances of WYSIWIG breaking down in traditional communication are rare and spark interest from researchers~\cite{hillberg2018what, cha2018complex}. In the case of immersive communication, however, messages are \emph{not} WYSIWIG. The sender cannot know, for example, exactly where or how an immersive message will appear in the recipient's physical world, which will affect the recipient's experience. As a result, reactions play an even more critical role in immersive communication than for traditional communication, so that the sender can see how the message appeared to the recipient and what the recipient's resulting experience was.

Our findings highlight design implications for reactions in immersive communication, including the role that authenticity should play in those reactions.

\subsubsection{Achieving both authenticity and control}
We used an automatic video capture as ARwand's form of reaction so that senders could see firsthand how their AR message was projected in the recipients' physical environment. When senders watched the reaction video that consisted of the AR content in the recipients' physical world, along with their reaction, they enjoyed it as a live recording and felt they were together even though they were apart. Some recipients, however, felt that they lacked control over the automatic recording, including the appearance of their surroundings in the video, even though they could reject the sending of the video in the end. Thus, future work should explore how to balance the authenticity of recording one's immediate reaction, with the desire for control over the actual reaction sent. To support better control, designers can incorporate ways to preview reaction messages or cancel the automatic recording of reactions. To respect recipients' comfort level, while preserving their initial reaction to an immersive message, designers might consider delaying the ``viewing'' of a received message to a later, more convenient time and place with an ability to manually initiate a reaction video. On a more fundamental level, future work can explore different reaction forms beyond full fledged videos, such as live photos, that capture the scene with AR message (WYSIWIG) and how the recipient actually felt---to help complete the loop of communication.  

\subsubsection{Control over reaction duration and battery life}
We implemented short reaction videos, primarily to preserve the battery life of the smartglasses. However, participants wanted to record longer reaction videos to continue the story or conversation that the sender created. This aligned with our findings that people use the immersive ability of AR to craft and tell stories in personal ways for their friends rather than simply to exchange information. Until smartglasses are free of battery life constraints, future designers can give control to the recipients to record longer videos with a constant update on battery life, leaving the decision of how to balance battery life with video duration to the user. Regardless, future immersive communication system designers can consider how recipients' reactions might be threaded into an ongoing, collaborative story that friends craft together by taking turns.

\subsection{Personalization and privacy challenges for AR immersive communication}
For immersive communication to be executed well for users, senders should be able to create personalized AR messages relevant to the recipient's context, and those AR messages must be delivered successfully (i.e., the trigger conditions must be successfully met at just the right place and time so that the recipient perceives it as highly immersive). We found that when triggers matched as the senders intended, it created a personalized and immersive experience for the recipient. \name did not share any of the recipients' information and was designed for close friends. Senders needed to have a detailed understanding of the recipients' whereabouts, e.g., when and where they are gonna be at a certain time, their physical space, e.g., artwork in their home, and their preferences to use triggers effectively. However, a recipient may change their plans and not share that information with the sender; thus, a sender might not always be aware of the recipient's context and triggers could fail to match conditions. If certain information about recipients were shared with the sender, it would increase the likelihood of a trigger match and benefit recipients' experience, but at the cost of their privacy. This tradeoff between effective immersive messaging and recipients' privacy leads to several future research challenges and design implications, which we describe below.

\subsubsection{Control over information sharing for recipients}
One challenge will be to understand how much information recipients are willing to share with others to have a better immersive messaging experience, how their preferences are shaped by context, and how recipients can be given additional control over how their information is shared. Prior work~\cite{liu2019animo, liu2021significant, griggio2019augmenting} has shown that people are more comfortable with sharing information about their activities with romantic partners and close friends. Likewise, we designed ARwand for close friends only since friends know enough details about each other to make triggers successful. However, people may have different preferences for sharing even amongst their friends, as evidenced by some participants who brought up privacy concerns. Thus, future work should explore methods that can preserve immersiveness while sharing less information, based on people's different relationships and sharing preferences.

An approach to address this privacy-functionality dilemma is to allow recipients to abstract their information. Location triggers, for example, could be based on labels such as ``Home,'' and ``Bedroom,'' instead of precise GPS locations. Immersive messages would then be delivered to predefined areas within those locations that the recipient configured. These could act as spatial mailboxes,  such as the recipient's nightstand for all messages sent to their ``Bedroom.''

On the other hand, some recipients may prefer more ways of sharing information to facilitate higher trigger success rates and more contextual AR messages. Future immersive communication systems could allow such recipients to share their calendar or schedule with others. Future systems could also allow recipients to upload scans of their room or office so that trusted senders can choose exactly where to place AR messages within their environment. Popular social apps such as Zenly~\cite{zenly}, Snapchat's Snap Map~\cite{snapmap}, and the Strava fitness app~\cite{strava} already explore ways for friends to share detailed information about their activities with others to spark rich conversations---where users can set automatic expiry duration to information sharing, limit the visibility to certain friend groups or can turn off the information sharing completely with a simple toggle button. The concept of immersive messaging shares many of the same design concerns and could potentially be integrated into activity-sharing platforms. By giving more control to the recipients, researchers can learn how people navigate the tradeoff between privacy and receiving contextually relevant messages in immersive communication and beyond.

\subsubsection{Control for senders}
Another research challenge will be understanding how to give senders greater control over how to personalize immersive messages for recipients, including giving senders appropriate fallback options in case a trigger condition cannot be met and the sender would still like the message delivered somehow. For instance, participants used time as a ``backup'' trigger. One method is to enable senders to show hints to the recipient after sending a message so that the recipient knows where or when to look for AR messages. Future AR communication systems could introduce new trigger types, such as activity or biosignals, and expand existing options, such as visual markers. An advanced version of a visual marker trigger can allow  senders to upload any picture as a visual marker, thereby increasing the chances of a trigger match and enabling greater control of contexts that can be selected as a trigger condition. If a sender is familiar with the recipient's favorite shirt, for example, they can simply find a picture of the brand's logo and set it as a visual marker so that the recipient receives the message when they wear it.

\ifnum\value{highlight}>0
}
\fi

\section{Limitations}
Our investigation was limited by the fact that we could not deploy \name in people's regular lives due to the ongoing COVID-19 pandemic. This meant that we could not observe context shifts (especially indoors) such as going to work, shopping, or spending time in cafés. We had to improvise to participants' everyday routine of mostly working from home and walking around their neighborhood to get some air.
Our results are also limited by participants not being able to borrow our smartglasses for an extended period of time. This is due to our prototype needing to be foolproof (sometimes requiring troubleshooting on-site) and smartglasses not yet being a mass market product.
\ifnum\value{revised}>0
{\color{blue}
\fi
Lastly, our study was limited due to the communication structure that participants experienced, where they could only play the role of either a sender or recipient. In their natural daily communication, people would typically act as both senders and recipients. Thus, future work should investigate how people interact when they can exchange messages back and forth. One next step could include having participants swap roles in order to understand their perceptions on both sides of communication.
\ifnum\value{revised}>0
}
\fi
Overall, despite these limitations, our research still contributes an 
\ifnum\value{highlight}>0
{\color{blue}
\fi
understanding people's perceptions of immersive interpersonal communication via AR messaging, and suggests important design and social implications that can benefit future work on immersive AR communication.


\ifnum\value{highlight}>0
}
\fi

\section{Conclusion}

In this paper, we explored how people use and perceive
\ifnum\value{highlight}>0 {\color{blue} \fi
AR messaging designed with the goal of investigating design factors and potential pitfalls of immersive AR communication.
\ifnum\value{highlight}>0
}
\fi
Our prototype system, \name, enables users to send \textit{AR messages} to a friend wearing smartglasses, directly augmenting their physical environment and creating experiences for them. The AR format changes the notion of what messages are. With AR, people can experience messages embedded in their reality.
In our user study with 24 participants, we observed that
\ifnum\value{highlight}>0
{\color{blue}
\fi
senders used AR messaging to create new stories based on shared memories and attempted to personalize messages according to the recipient's context. We observed that AR messages were enjoyable when appropriately embedded in the recipient's reality, but such immersion could sometimes be problematic when people use AR messaging in their daily lives. Additionally, recipients wanted to share their live experiences with AR messages with the senders but desired to have more control over this sharing interface. 
\ifnum\value{highlight}>0
}
\fi
Our exploratory research contributes to understanding of AR being a core medium of immersive interpersonal communication. 


\bibliographystyle{ACM-Reference-Format}
\bibliography{_paper}

\pagebreak
\appendix
\section{Appendix}

\begin{table*}[h]
    \centering
    \caption{Descriptive statistics for participants' overall experiences with \name.}
    \vspace{-0.8em}
    \resizebox{\textwidth}{!}{
    \begin{tabular}{l|l|l|l|l|l|l|}
         & \multicolumn{3}{c|}{\textbf{Senders}} & \multicolumn{3}{c|}{\textbf{Recipients}} \\
         & \textbf{Median} & \textbf{Mean} & \textbf{SD} & \textbf{Median} & \textbf{Mean} & \textbf{SD} \\
        \toprule
        \textbf{Overall experience --- Ratings (1: strongly disagree -- 7: strongly agree)} & & & & & & \\
        ``I feel that the system was \textbf{fun} to use'' & 6.5 & 6.42 & 0.67 & 6 & 5.67 & 1.67 \\
        ``I feel that the system was \textbf{easy} to use'' & 6.5 & 6.17 & 1.19 & 5 & 5 & 1.41 \\
        ``I feel that AR messages are \textbf{more effective} when they are relevant to the recipient’s situation'' & 6.5 & 6.5 & 0.52 & 6 & 6.17 & 0.83 \\
        ``I feel that AR messages \textbf{need to be relevant} to the recipient’s situation to be effective'' & 6 & 6.17 & 0.58 & 6 & 6.17 & 0.94 \\
        \makecell[l]{``I feel that AR messages are an \textbf{effective way} for me to communicate with my chosen friend\\ compared to other communication formats''} & 5.5 & 5.33 & 1.61 & 4.5 & 4.5 & 1.45 \\
        \bottomrule
    \end{tabular}
    }
    \label{tab:overall_exp}
\end{table*}

\begin{table*}[h]
    \centering
    \caption{Descriptive statistics for participants' experiences with creating or receiving AR messages.}
    \vspace{-0.8em}
    \resizebox{\textwidth}{!}{
    \begin{tabular}{l|l|l|l}
         & \textbf{Median} & \textbf{Mean} & \textbf{SD}  \\
        \toprule
        \textbf{Senders --- Ratings (1: strongly disagree -- 7: strongly agree)} & & & \\
        ``I \textbf{enjoyed sending} AR messages to my study partner'' & 6 & 6.08 & 0.79 \\
        ``I felt that I was able to \textbf{surprise} my study partner with my AR messages'' & 6 & 6.08 & 1 \\
        ``I felt that adding my \textbf{voice} to AR messages made our communication more effective`` & 6 & 6 & 1.04 \\
        ``I felt that I was able to \textbf{create experiences} for my study partner that are not possible in person'' & 5.5 & 5.83 & 1.11 \\
        ``I felt that I was \textbf{augmenting} my study partner’s reality when I sent AR messages'' & 6 & 6 & 0.95 \\
        ``I felt that I was \textbf{connected} to my study partner by sending AR messages'' & 6 & 5.17 & 1.64 \\
        ``I felt \textbf{comfortable} sending AR messages because the messages are for my study partner'' & 6 & 6.17 & 0.72 \\
        ``I felt that sending AR messages was \textbf{distracting} my study partner'' & 5 & 4.75 & 1.36 \\
        \midrule
        \textbf{Recipients --- Ratings (1: strongly disagree -- 7: strongly agree)} & & & \\
        ``I \textbf{enjoyed receiving} AR messages from my study partner'' & 6.5 & 6.17 & 1.4 \\
        ``I was \textbf{surprised} when I received AR messages from my study partner'' & 6 & 5.33 & 1.5 \\
        ``I felt that my study partner’s \textbf{voice} in the AR messages made our communication more effective'' & 6 & 5.83 & 1.27 \\
        ``I felt that my study partner was \textbf{augmenting} my reality via the AR messages'' & 6 & 5.08 & 1.31 \\
        ``I felt that I was \textbf{connected} to my study partner when I received AR messages'' & 6 & 5.58 & 1.16 \\
        ``I felt \textbf{comfortable} receiving AR messages because the messages are from my study partner'' & 5 & 5.17 & 1.4 \\
        ``I felt \textbf{distracted} by receiving AR messages from my study partner'' & 4.5 & 4.75 & 1.6 \\
        ``I enjoyed receiving AR messages containing \textbf{virtual objects}'' & 6 & 6 & 1.04 \\
        ``I enjoyed receiving AR messages containing my study partner’s \textbf{avatar}'' & 6 & 5.67 & 1.07 \\
    \end{tabular}
    }
    \label{tab:res_msg_exp}
\end{table*}

\begin{table*}[h]
    \centering
    \caption{Descriptive statistics for participants' experiences with sharing or viewing reaction videos.}
    \resizebox{\textwidth}{!}{
    \begin{tabular}{l|l|l|l}
         & \textbf{Median} & \textbf{Mean} & \textbf{SD}  \\
        \toprule
        \textbf{Recipients --- Ratings (1: strongly disagree -- 7: strongly agree)} & & & \\
        ``I \textbf{enjoyed} sharing my reaction with my study partner'' & 6 & 6.17 & 0.94 \\
        ``I felt \textbf{comfortable} sharing my reaction with my study partner'' & 6 & 6.17 & 0.72 \\
        ``I felt that I \textbf{had control} over sharing my reactions with my study partner'' & 5 & 4.25 & 1.96 \\
        ``I enjoyed sharing responses with my study partner without having to preview them first'' & 4.5 & 4 & 2.13 \\
        \midrule
        \textbf{Senders --- Ratings (1: strongly disagree -- 7: strongly agree)} & & & \\
        ``I \textbf{enjoyed} watching my study partner’s reactions to my AR messages'' & 6.5 & 6.25 & 0.87 \\
        ``I felt \textbf{comfortable} watching my study partner’s reaction video'' & 6 & 6.17 & 0.72 \\
        ``I felt \textbf{together} with my study partner when watching my study partner’s reaction videos & 6 & 5.92 & 0.79 \\
    \end{tabular}
    }
    \label{tab:res_video_exp}
\end{table*}

\begin{table}[h]
    \centering
    \caption{Message count and delivery rate for each participant. For senders, the count shows how many messages they scheduled using each of single triggers or a compound trigger with either the Specific (``AND'') mode or the Flexible (``OR'') mode. For Wearers, the count shows how many messages they eventually received from their sender.}
    \resizebox{\textwidth}{!}{
    \begin{tabular}{l|l|l|l|l|l||l|l|l|l|l|l|l|l|l|l|l}
         & \multicolumn{3}{l|}{Single trigger} & \multicolumn{2}{l||}{Compound trigger} & & \multicolumn{6}{l|}{Single trigger} & \multicolumn{4}{l}{Compound trigger} \\
        \hline
        \multirow{2}{*}{\textbf{PID}} & \textbf{Location} & \textbf{Time} & \textbf{Marker} & \textbf{Specific} & \textbf{Flexible} & \multirow{2}{*}{\textbf{PID}} & \multicolumn{2}{l|}{\textbf{Location}} & \multicolumn{2}{l|}{\textbf{Time}} & \multicolumn{2}{l|}{\textbf{Marker}} & \multicolumn{2}{l|}{\textbf{Specific}}  & \multicolumn{2}{l}{\textbf{Flexible}} \\
         & Count & Count & Count & Count & Count & & Count & Rate & Count & Rate & Count & Rate & Count & Rate & Count & Rate \\
        \hline
        S1 & 1 & 3 & 3 & 0 & 0 & W1 & 1 & \textbf{100\%} & 0 & 0\% & 3 & \textbf{100\%} & 0 & \textit{N/A} & 0 & \textit{N/A} \\
        S2 & 2 & 3 & 2 & 1 & 0 & W2 & 0 & 0\% & 1 & 33\% & 2 & \textbf{100\%} & 0 & 0\% & 0 & \textit{N/A} \\
        S3 & 1 & 1 & 1 & 1 & 0 & W3 & 1 & \textbf{100\%} & 1 & \textbf{100\%} & 0 & 0\% & 0 & 0\% & 0 & \textit{N/A}\\
        S4 & 5 & 1 & 2 & 1 & 0 & W4 & 5 & \textbf{100\%} & 1 & \textbf{100\%} & 2 & \textbf{100\%} & 1 & \textbf{100\%} & 0 & \textit{N/A} \\
        S5 & 0 & 2 & 0 & 6 & 1 & W5 & 0 & \textit{N/A} & 1 & 50\% & 0 & \textit{N/A} & 0 & 0\% & 0 & 0\% \\
        S6 & 3 & \textbf{5} & 1 & 0 & \textbf{8} & W6 & 1 & 33\% & 0 & 0\% & 1 & \textbf{100\%} & 0 & \textit{N/A} & 5 & 63\% \\
        S7 & 1 & 0 & \textbf{7} & \textbf{8} & 2 & W7 & 0 & 0\% & 0 & \textit{N/A} & 1 & 14\% & 0 & 0\% & 1 & 50\% \\
        S8 & 1 & 1 & 1 & 1 & 6 & W8 & 1 & \textbf{100\%} & 1 & \textbf{100\%} & 1 & \textbf{100\%} & 0 & 0\% & 6 & \textbf{100\%} \\
        S9 & \textbf{6} & 1 & 4 & 0 & 2 & W9 & 3 & 50\% & 0 & 0\% & 2 & 50\% & 0 & \textit{N/A} & 1 & 50\% \\
        S10 & 4 & 2 & 2 & 0 & 0 & W10 & 3 & 75\% & 0 & 0\% & 2 & \textbf{100\%} & 0 & \textit{N/A} & 0 & \textit{N/A} \\
        S11 & 1 & 1 & 1 & 0 & 4 & W11 & 1 & \textbf{100\%} & 0 & 0\% & 1 & \textbf{100\%} & 0 & \textit{N/A} & 4 & \textbf{100\%} \\
        S12 & 2 & 1 & 5 & 1 & 0 & W12 & 2 & \textbf{100\%} & 1 & \textbf{100\%} & 5 & \textbf{100\%} & 0 & 0\% & 0 & \textit{N/A} \\
        \midrule
        \textit{Median} & \textit{1.5} & \textit{1} & \textit{2} & \textit{1} & \textit{0.5} & \textit{Median} & \textit{1} & \textit{100\%} & \textit{0.5} & \textit{33\%} & \textit{1.5} & \textit{100\%} & \textit{0} & \textit{0\%} & \textit{0} & \textit{57\%} \\
        \textit{Mean} & \textit{2.25} & \textit{1.75} & \textit{2.42} &  \textit{1.58} & \textit{1.92} & \textit{Mean} & \textit{1.5} & \textit{68.9\%} & \textit{0.5} & \textit{43.9\%} & \textit{1.67} & \textit{78.5\%} & \textit{0.08} & \textit{14.3\%} & \textit{1.42} & \textit{60.5\%} \\
        \textit{SD} & \textit{1.86} & \textit{1.36} & \textit{2.02} & \textit{2.61} & \textit{2.71} & \textit{SD} & \textit{1.51} & \textit{41.2} & \textit{0.52} & \textit{47.3} & \textit{1.37} & \textit{38.5} & \textit{0.29} & \textit{37.8} & \textit{2.23} & \textit{37.4} \\
    \end{tabular}
    }
    \label{tab:stat_msg_schedule}
\end{table}

\begin{figure}
    \centering
    \begin{subfigure}{0.3\textwidth}
        \includegraphics[width=\textwidth]{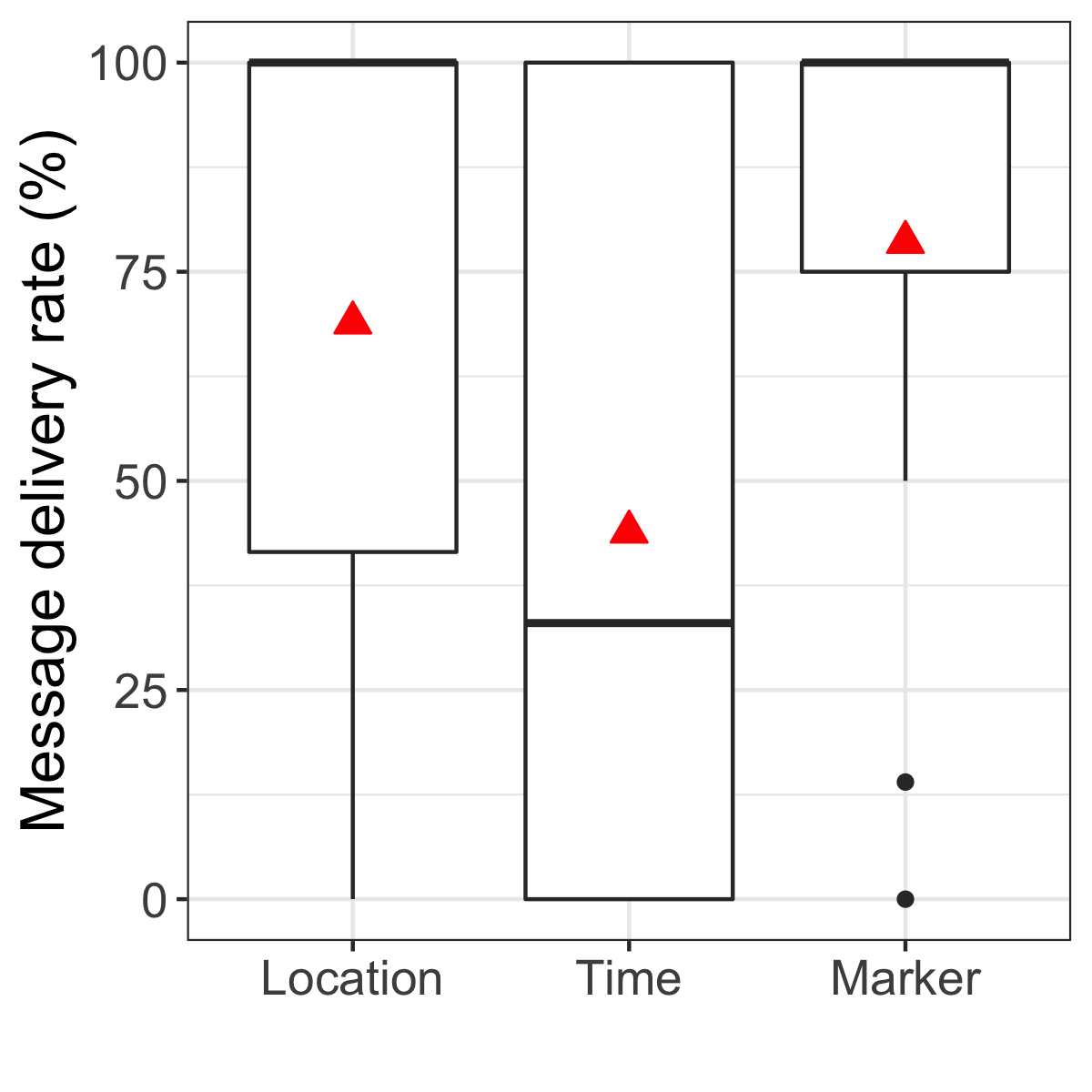}
        \caption{Single trigger}
        \label{fig:single_trigger_rate}
        \Description[The box plot describing the message delivery rate in percent for each single trigger use: Location, Time, and Visual marker]{Location trigger: median=100\%, mean=68.9\%, sd=41.2; Time trigger: median=33\%, mean=43.9\%, sd=47.3; and Visual marker trigger: median=100\%, mean=78.5\%, sd=38.5}
    \end{subfigure}
    \qquad
    \begin{subfigure}{0.3\textwidth}
        \includegraphics[width=\textwidth]{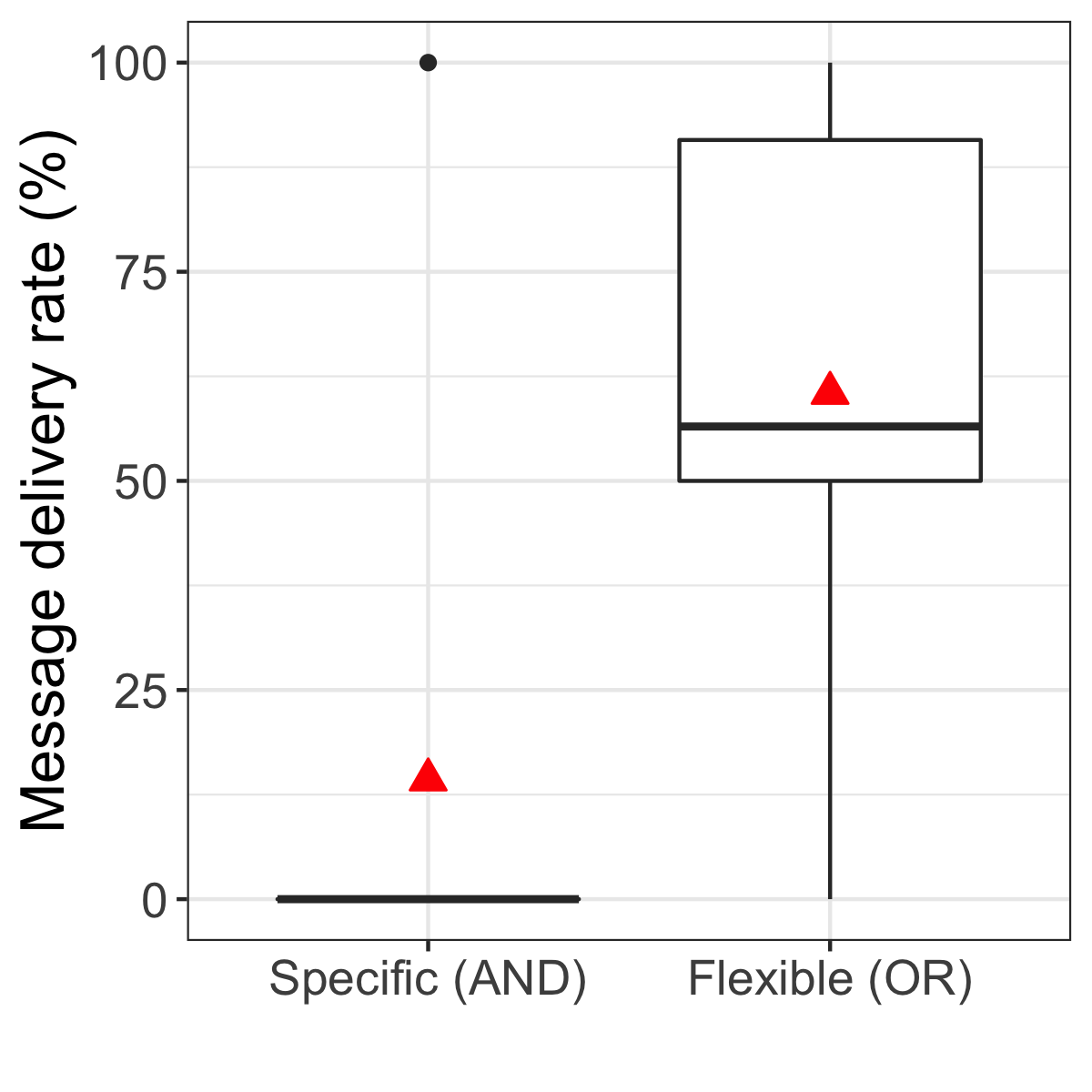}
        \caption{Compound trigger}
        \label{fig:comp_trigger_rate}
        \Description[The box plot describing the message delivery rate in percent for compound trigger use: Specific (AND) and Flexible (OR)]{Specific (AND) option: median=0\%, mean=14.3\%, sd=37.8; and Flexible (OR) option: median=56.5\%, mean=60.5\%, sd=37.4}
    \end{subfigure}
    \caption{Success rate for scheduled messages. (a) Visual marker triggers were successful most often, and time triggers had the least success. Time triggers failed whenever they were set for times that Wearers were not wearing their smartglasses. (b) The ``Flexible'' (OR) option for compound triggers increased the success rate dramatically compared to the ``Specific'' (AND) option. Table~\ref{tab:stat_msg_schedule} in our appendix shows detailed statistics for scheduled messages.}
    \label{fig:msg_delivery}
\end{figure}

\end{document}